\documentclass[10pt,a4paper,twocolumn]{article}
\usepackage[T1]{fontenc}
\usepackage[utf8]{inputenc}
\usepackage{lmodern}
\usepackage[margin=20mm]{geometry}
\usepackage{amsmath,amssymb}
\usepackage{microtype}
\usepackage[hidelinks]{hyperref}
\hypersetup{pdftitle={pyFDN: A Python Toolbox for Feedback Delay Network Design, Analysis, and Optimization},pdfauthor={Sebastian J. Schlecht, Facundo Franchino, Jeremy B. Bai}}
\newcommand{\pyfdn}{\texttt{pyFDN}}
\title{\vspace{-8mm}pyFDN:\\A Python Toolbox for Feedback Delay Network\\Design, Analysis, and Optimization}
\author{%
Sebastian J. Schlecht\textsuperscript{1}\quad
Facundo Franchino\textsuperscript{2,3}\quad
Jeremy B. Bai\textsuperscript{1}\\[3pt]
\small \textsuperscript{1}Friedrich-Alexander-Universit\"at Erlangen-N\"urnberg,\\[-1pt]
\small Multimedia Communications and Signal Processing, Germany\\
\small \textsuperscript{2}Research Laboratory of Electronics, Massachusetts Institute of Technology, USA\\
\small \textsuperscript{3}School of Physics, Engineering and Technology, University of York, UK}
\date{}
\begin{document}
\raggedbottom
\maketitle
\begin{abstract}
Feedback delay networks (FDNs) provide a compact representation of recursive audio systems for artificial reverberation and decorrelation. Their practical use spans structured matrix design, acoustic decay shaping, numerical simulation, and modal analysis, increasingly combined with differentiable optimization. We present \pyfdn{}, an open-source Python toolbox that brings these operations into a common workflow. The library combines numerical design and analysis routines with block-based time-domain processing and integration with FLAMO for frequency-domain differentiable processing. Explicit network builds and JSON presets connect design, optimization, rendering, and reuse. This paper describes the software architecture, representative capabilities, and reproducibility mechanisms of version 0.5.0, providing a reference for research and teaching that use the toolbox.
\end{abstract}

\section{Motivation and scope}
Feedback delay networks (FDNs) combine delay lines with a feedback matrix. They are widely used for artificial reverberation and decorrelation, and provide a compact model for studying spectral coloration, echo density, and modal decay. A typical experiment requires several views of the same network. One may construct its matrices, listen to its response, examine its modes, and adjust its parameters against a measurement.

The legacy Feedback Delay Network Toolbox (FDNTB) collected tools for FDN design and analysis in MATLAB~\cite{fdntb}. \pyfdn{} brings related methods into Python and connects them to FLAMO's differentiable audio models~\cite{flamo}. The software contribution comprises three connected workflows: construction and time-domain rendering (Sections~2--3), representation conversion and modal analysis (Section~3), and differentiable optimization with numerical export (Section~4). These workflows integrate established methods; this overview makes no comparative performance or perceptual claim.

\section{Network representation}
For a linear time-invariant (LTI) FDN with $N$ delay lines of positive integer lengths $m_i$ and a static feedback matrix, the unfiltered transfer matrix is
\begin{equation*}
\mathbf{H}(z)=\mathbf{C}\bigl[\mathbf{\Delta}(z)-\mathbf{A}\bigr]^{-1}\mathbf{B}+\mathbf{D},
\end{equation*}
where $\mathbf{\Delta}(z)=\operatorname{diag}(z^{m_1},\ldots,z^{m_N})$. The feedback matrix is $\mathbf{A}\in\mathbb{R}^{N\times N}$, while $\mathbf{B}\in\mathbb{R}^{N\times K}$ maps $K$ inputs to the delay lines and $\mathbf{C}\in\mathbb{R}^{L\times N}$ maps their outputs to $L$ output channels. The matrix $\mathbf{D}\in\mathbb{R}^{L\times K}$ supplies the direct path. An ordinary state-space realization has $\sum_i m_i$ states. Both time-domain and FLAMO processing also support finite impulse response (FIR) feedback matrices and additional filters.

The \texttt{generate}, \texttt{translate}, \texttt{td}, \texttt{eq}, and \texttt{train} modules cover construction, conversion, time-domain processing, equalization, and optimization. NumPy and SciPy support numerical routines, and FLAMO and PyTorch support differentiable models.

An \texttt{FDNBuild} holds the matrices, delays, sample rate, and optional second-order-section (SOS) filter banks. Filters may follow the delays, the feedback matrix, or the wet output. The direct path bypasses the output filter. An \texttt{FDNPreset} adds metadata and optional design records. Its numerical build determines the saved network, which can be rendered without repeating the design or optimization. The JSON format supports static feedback matrices and SOS filter banks.

\section{Design and analysis}
\paragraph{Network construction.}
Generators provide random orthogonal and Householder matrices, Kronecker constructions, and cascaded paraunitary FIR matrices. Allpass FDN constructions and completion routines implement established delay-network designs~\cite{allpass}. Delay sampling and matrix galleries support systematic experiments. Examples include scattering delay networks~\cite{sdn}, grouped FDNs for coupled rooms~\cite{gfdn}, time-varying feedback~\cite{timevarying}, and decorrelation~\cite{decorrelation}.

\paragraph{Acoustic decay and equalization.}
For an unfiltered lossless prototype with static feedback, a uniform modal decay time $T_{60}$ (60\,dB in amplitude) is obtained by applying the amplitude gain
\begin{equation*}
g_i=10^{-3m_i/(f_sT_{60})}
\end{equation*}
to each delay line, where $f_s$ is the sample rate and $T_{60}$ is measured in seconds. One-pole, first-order shelving, and graphic equalization filters approximate frequency-dependent decay targets. Output equalization shapes the wet response outside the feedback loop. Acoustic diagnostics include energy-decay curves, bandwise reverberation-time estimates, and echo density.

\paragraph{Time-domain rendering.}
\texttt{process\_fdn} renders a build, and \texttt{process\_dss} accepts raw matrices, delays, and filter hooks. Each call constructs a fresh processing graph. For continuous processing, \texttt{build\_to\_td} and \texttt{dss\_to\_td} return graphs that retain state across blocks. The \texttt{td} operators and their series, parallel, and recursive connections also support time-varying and nonlinear structures.

\paragraph{Alternative representations.}
Unfiltered delay state space can be converted to transfer functions, impulse responses, and ordinary state space. The ordinary state-space converter requires a single input and output and $m_i\geq3$. For static feedback matrices, pole--residue analysis offers eigenvalue, characteristic-polynomial root, and Ehrlich--Aberth methods. The FLAMO solver also supports SOS loop filters within a single, unnested recursion. Poles describe modal frequency and decay, while residues describe excitation at the chosen inputs and outputs~\cite{modal}.

\section{Differentiable optimization}
Training constructs a FLAMO model, minimizes a weighted sum of losses, and extracts a numerical build. The interface can train the feedback matrix, input and output gains, and direct path while keeping delays fixed. The feedback matrix can be constrained to the special orthogonal group $\mathrm{SO}(N)$ or trained without that constraint. Dedicated filter modules expose reverberation times and equalization gains as trainable parameters. SOS coefficients loaded from a build remain fixed by default. Preset design records can recover filter targets when they reproduce the saved coefficients.

Losses compare magnitudes, spectrograms, impulse responses, and energy decays, with optional parameter penalties. Examples cover colorless reverberation~\cite{colorless}, fitting filters to room impulse responses~\cite{filters}, and coupled-room modeling~\cite{gfdn}. A designed network can initialize an optimization, and \texttt{extract\_build} returns its learned matrices and supported filter banks for time-domain rendering and preset storage. A separate integration example uses ADAC to compile a FLAMO model to FAUST~\cite{adac}.

Frequency-domain synthesis depends on the transform length and control of time aliasing. FLAMO's alias-decay setting suppresses wraparound through a temporary damping envelope that is removed from the synthesized response and leaves the extracted build unchanged. Removing this envelope can amplify round-off error, so its setting must be chosen together with numerical precision. Long and lossless decays require particular care. Modal analysis depends on numerical conditioning and solver settings. Orthogonal and paraunitary constructions supply lossless prototypes. Stability of the complete network also depends on its filters and parameter constraints.

\section{Availability}
\pyfdn{} is available from PyPI (\texttt{pip install pyFDN}) and its public repository\footnote{\url{https://github.com/artificial-audio/pyFDN}}. Documentation provides an API reference and executable examples\footnote{\url{https://artificial-audio.github.io/pyFDN/}}. Tests cover construction, processing, conversions, presets, and training, including comparisons between time-domain, frequency-domain, and state-space responses.

This overview describes version 0.5.0 at source revision \texttt{3732d6b0d334}. The code is MIT-licensed. Bundled resources retain their documented terms, including the noncommercial restriction on the Pori concert-hall impulse response. Reproducible experiments should record the software revision, preset, random seeds, dependency versions, and numerical settings, including transform length, alias decay, precision, and solver or training settings. The preset stores the sample rate and realized network parameters.

\section{Conclusion}
\pyfdn{} connects FDN construction, rendering, analysis, and differentiable optimization through shared representations and saved numerical builds. It provides reusable implementations and examples for research and teaching, with explicit limits on serialization, conversion, and model structure.

\end{document}